\documentclass[journal=mamobx,manuscript=article]{achemso}
\usepackage{xcolor}
\usepackage{multirow}
\usepackage{makecell}
\usepackage{graphicx}
\usepackage{pdfpages}
\usepackage{caption}
\usepackage{subcaption}
\SectionNumbersOn        
\usepackage[version=3]{mhchem} 

\newcommand{\Will}[1]{\textcolor{black}{#1}}

\author{William S. Fall}
\email{william.fall@cnrs.fr}
\affiliation{Laboratoire de Physique des Solides - UMR 8502,
CNRS, Universit\'e Paris-Saclay, 91405 Orsay, France}

\author{J\"org Baschnagel}
\affiliation{Institut Charles Sadron, Universit\'e de Strasbourg \& CNRS, 23 rue du Loess, 67034 Strasbourg Cedex, France.}

\author{Hendrik Meyer}
\affiliation{Institut Charles Sadron, Universit\'e de Strasbourg \& CNRS, 23 rue du Loess, 67034 Strasbourg Cedex, France.}

\title[An \textsf{achemso} demo]
  {Branches, Tie Chains and Entanglements in Bimodal Polyethylene Single Crystals under Uniaxial Tensile Strain}

\abbreviations{IR,NMR,UV}
\keywords{American Chemical Society, \LaTeX}

\begin{document}

\begin{abstract}

Using coarse-grained molecular dynamics simulations and a united-monomer model of PE, single well-aligned multi-lamella PE crystals grown in previous work [ACS Macro Letters \textbf{12}, 808 (2023)] 
are deformed uniaxially to mimic tensile testing. During deformation, the crystallinity, tie-chain segments, entanglements and folds are monitored and correlated with the stress-strain behaviour and mechanical properties. At small strains, the single well-aligned PE crystals reveal a larger Young modulus when the deformation direction is perpendicular to the global stem direction. At large strains, the memory of the initial topology plays little role in the mechanical response and is observed to be completely destroyed. Short chain branching appears to suppress disentanglement and shear induced alignment of the chains during deformation. As a result, tie-chains and entanglements persist in branched systems and the peak stress at failure is found to be proportional to the change in number of tie-chains from the beginning of the brittle break to its end. Our findings suggest the remarkable mechanical properties of bimodal branched PE result directly from tie-chains, with entanglements playing a secondary role in the mechanical response. 

\end{abstract}

\section{Introduction}

High-pressure plastic gas or water pipelines must be long-lasting, provide 
provide resistance against slow crack growth and rapid crack propagation and withstand extreme temperature and pressure fluctuations under operating conditions \cite{garcia2011effects,hedenqvist1996fracture,hubert2001physical,hubert2002physical,clutton1998slow,ZHA2022104663}. The best performing plastics which fulfill all these requirements are bimodal polyethylene (PE) resins\cite{deslauriers2005comparative,bohm1992high,knuuttila2004advanced}, specifically those with short chain branches concentrated in the high molecular weight (HMW) portion \cite{moreno2017production}. Industrial manufacturers of such high-performing plastics regard their synthesis and polymerisation steps to be highly sensitive and are often kept confidential \cite{olhostprivatecomm}. In bimodal PE, the high elastic modulus and good processability of the low molecular weight (LMW) portion may be combined with the mechanical properties, such as flexibility and resistance to stress or slow cracking, of the HMW portion \cite{sun2011fracture,soares2000environmental,qin2022environmental}. Control over the HMW/LMW fractions and distribution of short/long chain branches (SCBs/LCBs) within the MW distribution is however extremely challenging \cite{berthold1996advanced}. Moreover relating the MW distribution and the placement or concentration of SCBs/LCBs to mechanical properties is often hampered by the inherent polydispersity and different crystallisation protocols \cite{ungar2001learning}. This makes it difficult to relate the molecular architecture to large-scale semi-crystalline properties, such as the concentration of tie-chains and entanglements and ultimately how they are correlated with mechanical properties \cite{seguela2005critical}. The role of tie chains and entanglements on the mechanical properties of PE materials has been a subject of controversy for several decades because direct imaging is experimentally challenging \cite{https://doi.org/10.1002/mame.202100536,mcdermott2020measuring}. 

It is the interconnected nature of crystallites through the amorphous region, arising from tie-chains and entanglements, that is responsible for the unique mechanical properties of semi-crystalline PE. Without these interconnections, the crystal planes formed by polymer chains are only weakly bound by van der Walls interactions, meaning they can slide past or split apart from one another under small tensile strains \cite{}. This results in very brittle materials which possess poor mechanical properties. In general, PE materials are susceptible to slow crack growth, which leads to brittle failure at stress levels far below the yield point \cite{brown1992fundamental,kausch1978deformation}. It has been postulated that this arises due poorly interconnected crystallites and small number of interconnects which gradually lose mechanical efficiency under an applied load, either by chain scission or disentanglement \cite{deblieck2011failure}. Branched bimodal PE, however, demonstrates slow crack growth resistance which is believed to arise from an increased number of tie-chains \cite{bohm1992high} and as such it has been the subject of many experimental studies to determine its origin and optimise its properties \cite{brana1992morphology,garcia2011effects,bohm1992high,hedenqvist1996fracture,hubert2001physical,sun2011fracture,deslauriers2018chemometric,long2023molecular,bashir1986stiff}. However, it is not possible experimentally to directly measure the number of interconnects between crystallites. Experimental studies often rely on theoretical models, such as the Huang and Brown tie chain model, to approximate the number of tie chains \cite{huang1988effect}. Combined with other data, such as the molar mass and branch distributions, it is currently one of the best metrics, to correlate tie-chains with fracture toughness, available \cite{deslauriers2018chemometric}.    

Over the last few decades, simulations have emerged as a new tool to understand structure-property relationships in polymer materials, where quantities such as the molecular weight distribution or branch, tie-chain and entanglement content are known precisely. To date, simulations of PE have addressed many aspects including nanostructures formed during crystallisation \cite{LuoSommer:PRL2014,ko2004characterization,yi2013molecular,sommer2010molecular,morthomas2017crystallization,ramos2015molecular,meyer2001formation,meyer2002formation,hall2019coarse,hall2019divining,hall2021chain,hall2020monodisperse,gee2006atomistic,chen2022molecular} and mechanical deformation properties \cite{men2003role,yeh2017molecular,yeh2015mechanical,kim2014plastic,jabbari2015correlation,higuchi2017deformation,ranganathan2020atomistic,grommes2021investigation,yoshida2023effects,zhang2022molecular,higuchireversibility} or indeed morphological features such as tie-chains and entanglements in the final semi-crystalline form \cite{LuoSommer:PRL2014,yeh2017molecular,kumar2017effect,zhai2019disentangling}. Some simulation studies have studied bimodal PE materials \cite{triandafilidi2016molecular,luo2016role,jiang2015understanding,hu2018effect,moyassari2019molecular,moyassari2019molecular2,zhai2019disentangling,cao2021molecular,zhai2023enhanced} or systems containing short chain branches \cite{moyassari2019molecular,kumar2017effect,zhang2018direct,zhai2019crystallization,hu2018effect,hu2019dominant,ranganathan2020atomistic,sanmartin2012following,sanmartin2014strong,zhang2006differences,fall2022role,fall2023molecular,zhang2021modeling}. The main focus of many of these studies is often on the crystallisation step and only a few groups have thus far addressed mechanical properties in bimodal PE \cite{moyassari2019molecular,song2021molecular} or branched bimodal PE \cite{moyassari2019molecular2}. In addition, a continuous-cooling crystallisation protocol is usually employed, resulting in multi-domain systems which are poorly oriented \cite{gee2006atomistic,fall2023molecular}. Addressing the direction dependent deformation properties of single PE crystals has therefore remained relatively unexplored by simulations. In our previous study, well-aligned multi-lamellar single PE crystals were grown from the melt by applying a self-seeding protocol to bimodal branched PE systems \cite{fall2023molecular}. Such well-aligned systems are the first of their kind and allow us to uniquely address the direction dependent deformation properties of single PE crystals in bimodal branched PE and compare to  multi-domain systems grown by continuous-cooling. We also compare the branched systems with bimodal systems without chain branching. Short chain branches are found to have an important impact on the ultimate mechanical strength and failure behavior of the polymer crystals. We find that this impact strongly correlates with the evolution of the number of tie-chains between crystalline domains upon deformation, while the evolution of chain entanglements appears to be of secondary importance only.

Our article is structured as follows. Section~\ref{sec:model} presents the simulation model and simulation methods, including the deformation protocol and the analysis of tie chains and entanglements. Section~\ref{sec:results} discusses the results obtained for the different systems first for small strains in the Young's modulus region and then later at larger strains where we correlate the stress-strain behavior with morphological features. This includes folds, tie-chains, entanglements and the deformation-induced crystalline of the chains in the semi-crystalline microstructure. In Section \ref{sec:conclusions} the main findings are summarised and related to the known mechanical properties of real bimodal PE systems seen experimentally.

\section{Model, Methods and Deformation Protocol}
\label{sec:model}

The series of polymer crystals, used for tensile testing in this work, have been reported in a previous study where multi-lamellar single PE crystals were grown from the melt using self-seeding \cite{fall2023molecular}. All systems have been crystallised from a fully amorphous polymer melt and short chain branches are included using the procedure outlined previously in Ref.\citenum{fall2022role}. A united-monomer model of PE is utilised, where a chemical monomer is represented by a single coarse-grained (CG) bead in the simulation. Three systems are studied. First, a linear bidisperse polymer melt without chain branching, containing 3,840 200-bead chains (768,000 united-monomers) and 384 2,000-bead chains (768,000 united-monomers) which makes a total of 1,536,000 united-monomers. Moreover, starting from this linear bidisperse melt, either 10 or 20 short chain branches (SCBs) are introduced into the HMW portion, on the long chains only, providing two additional systems with 1,543,680 and 1,551,360 united-monomers, respectively. 

\Will{The united-monomer model used to grow the polymer crystals incorporates both stretching and bending valence terms as well as soft non-bonded purely repulsive interactions in the form of a Weeks-Chandler-Anderson (WCA) potential \cite{fall2022role,fall2023molecular}. In previous studies it has been demonstrated sucessfully that a purely repulsive potential coupled with a high pressure is sufficient to induce crystallisation. This approximation is beneficial since simulations are less computationally demanding with a smaller cutoff. Before deformation can be carried out however, the cutoff of the WCA potential is modified to include the attractive tail since this high pressure approximation is no longer valid in the non-equilibrium scenario. The long-ranged van der Waals attractive interactions are therefore recaptured by the force-field with the inclusion of some of the attractive tail. Both potentials have the same form and differ only in their respective cutoffs at large-distances and are referred to as WCA (purely repulsive) or LJ otherwise. See Table \ref{tbl:params} for a summary of potential forms and parameters originally outlined in [\!\!\citenum{fall2022role}].} 

\begin{table*}
\caption{\Will{Summary of the potential forms and parameter values for the united-monomer model of PE \cite{fall2022role,fall2023molecular}. Note the WCA potential form is used during crystallisation and the LJ (with attractive tail) during deformation. Both potential forms are otherwise identical and differ only in their respective cutoffs at large distances. }}
\label{tbl:params}
\footnotesize
\setlength{\tabcolsep}{1pt}
\begin{tabular}{c|c|c|c}
& Interaction & Form & Parameters \\
\hline
\hline
\multirow{3}{*}{Bonds} & \multirow{3}{*}{\makecell{(C\textsubscript{2}H\textsubscript{4})-(C\textsubscript{2}H\textsubscript{4})\\
}} & \multirow{3}{*}{$U_{\mathrm{bond}}(l)\texttt{=}\frac{1}{2}k_{\mathrm{bond}}(l-l_{0})^{2}$} & \multirow{3}{*}{\makecell{$k_{\mathrm{bond}}\texttt{=}53.69$ (kcal/mol/\AA\textsuperscript{2})\\ $l_{0}\texttt{=}2.225$ (\AA)}}  \\
& & & \\
& & & \\
\hline
\multirow{1}{*}{Angles} & (C\textsubscript{2}H\textsubscript{4})-(C\textsubscript{2}H\textsubscript{4})-(C\textsubscript{2}H\textsubscript{4}) & Tabulated &  See [\!\!\citenum{fall2022role}] or SI. \\
\hline
\multirow{4}{*}{\makecell{Non-Bonded}} & \makecell{(C\textsubscript{2}H\textsubscript{4})$\leftrightarrow$(C\textsubscript{2}H\textsubscript{4})\\ Crystallisation (WCA) \\ Repulsive} & \multirow{4}{*}{\makecell{\(\scalebox{1.2}{$\,U_{\mathrm{WCA/LJ}}^{({9\texttt{-}6})}\texttt{=}4\epsilon_{0}\bigg[\Big(\frac{\sigma_{0}}{r}\Big)^{9}\texttt{-}\Big(\frac{\sigma_{0}}{r}\Big)^{6}\bigg]$}\)}} & \makecell{$\epsilon_{0}\texttt{=}0.348$ (kcal/mol)\\$\sigma_{0}\texttt{=}4.45$ (\AA) \\ $r_{\mathrm{c}}^{(\mathrm{WCA})}=(3/2)^{\frac{1}{3}}\sigma_{0}$}   \\
\cline{2-2} 
\cline{4-4}
& \makecell{(C\textsubscript{2}H\textsubscript{4})$\leftrightarrow$(C\textsubscript{2}H\textsubscript{4})\\ Deformation (LJ) \\ Attractive} & & \makecell{$\epsilon_{0}\texttt{=}0.348$ (kcal/mol)\\$\sigma_{0}\texttt{=}4.45$ (\AA) \\ $r_{\mathrm{c}}^{(\mathrm{LJ})}=2.0\sigma_{0}$} \\
\hline
\end{tabular}%
\end{table*} 

To facilitate tensile testing, a sufficiently long attractive tail of the WCA potential must be included, otherwise known at a Lennard-Jones (LJ) potential. The cutoff of the potential was therefore modified to 2.0$\sigma$ and a new equilibrium pressure applied $P=1.8$ ($k_\mathrm{B} T_0/\sigma^3$), obtained from crystallisation studies with attractive interactions included. The semi-crystalline multi-lamellar PE crystals are then subject to equilibration runs in the NPT ensemble at constant pressure $P$ via a Nos\'e-Hoover thermostat and an anisotropic barostat, with coupling constants $T_{\mathrm{damp}}=2.0$  ($\tau$) and $P_{\mathrm{damp}}=20.0$ ($\tau$). Note the barostat is coupled only to the two box dimensions not under deformation independently, corresponding to uniaxial tensile strain conditions. \Will{Thus each direction has an independent equilibration to relax internal stresses before deformation begins.} \Will{We always use the integration timestep 0.005~$\tau$, where the LJ-time unit $\tau = \sqrt{m\sigma^2/k_\mathrm{B}T_{0}}$ corresponds to 2.7~ps with $m=27.3881$~g/mol, $T_{0}=500$K and $\sigma=0.5$nm, see [\!\!\citenum{fall2022role}] for the derivation of SI units. Temperature is held fixed at $T=300$~K $= 27^\circ$C ($T=0.6$ in reduced units) and equilibration runs are performed for 10,000 $\tau$ to remove any anisotropy in the pressure tensor prior to tensile testing. }

\Will{During production runs each dimension of the box is deformed independently to assess the dependence on direction and two strain rates of 10$^{3}$$s^{-1}$ and 10$^4$$s^{-1}$ are considered to understand how the stress-strain behaviour depends on the rate of deformation. The strain rate may be defined as $\varepsilon=\text{d} \lambda(t) / \text{d} t$ where $\lambda(t) = L(t) / L(t=0)$ with $L(t)$ being the linear size of the simulation box at time $t$. True strain, corresponds to the instantaneous force per current deformed area and is defined as $\sigma_{\mathrm{t}}=F_{i}(t)/A(t)$, where $F_{i}(t)$ is the component of the stress tensor in the deformation direction $i$ and $A(t)$ is the current area of the cross-section of the simulation box perpendicular to the deformation direction at time $t$. The stress in direction $i$ ($\sigma_i$) is $\sigma_{i} = - P_{i} + P_0$ where $P_0 = 1.8$ ($k_\mathrm{B} T_0/\sigma^3$), is the equilibrium pressure. In this way, the undeformed configuration is stress-free and we have in the linear regime $\sigma_{i} = E \varepsilon_{i} \simeq E (1 + \varepsilon_{i}) = E [L_{i}(t)/ L_{i}(t=0)]$ where $\varepsilon_i = [L_{i}(t) - L_{i}(t=0)]/L_{i}(t=0)$. The tie-chain and entanglement content are then continuously monitored throughout.} 

\subsection{Tie Chain Segment Analysis}

To identify a tie-chain segment, the local (second Legendre polynomial) $P_{2}$ order parameter is leveraged, as commonly used in simulations of polymer crystallisation \cite{meyer2002formation,meyer2001formation,hall2019coarse,hall2019divining,moyassari2019molecular,moyassari2019molecular2,zhang2021quasi,zhang2021roughening,yamamoto2019molecular,fall2022role,fall2023molecular}. This allows a local value of crystallinity to be assigned to each CG bead along the polymer chain. For the sake of clarity, we outline this approach here. The local ordering of beads within a given cutoff distance is determined using the angle $\theta_i$ of a given bead $i$ with all neighbouring beads inside a given cutoff $r_\text{c}$ according to $P_{2}(i) = \langle (3\cos^{2}{\theta_i}-1)/2\rangle$ where the average is carried out over all neighbours of $i$ within $r_\text{c}$. \Will{The direction of a given bead is the vector spanning the nearest neighbouring bead positions $p_{i\pm1}$ such that $\vec{v_{i}}=p_{i-1}-p_{i+1}$ with the exception of the chains ends where the bond vector is used to assign a direction i.e. $\vec{v_{i}}=p_{i}-p_{i\pm1}$.} The second minimum of the radial distribution function is used as the cutoff distance ($\sim 1.6\sigma$) corresponding to the next nearest neighbour, see Ref.~\citenum{fall2022role}. \Will{Beads with $P_{2} = 0.85$ are considered crystalline because this value corresponds to the upturn of the \textit{trans-trans} minimum of the bond angle distribution, corresponding to an minimum possible average angle of approximately 145$^\circ$ of a bead with its neighbours to be considered crystalline}. \Will{With this information, all unbroken contiguous chain segments in the system can be determined. Tie-chains can then be identified by first traversing each chain, finding unbroken contiguous segments of crystalline beads and assigning them to an unbroken stem.} Then, each stem is assigned a directional vector spanning the first and last beads along the stem. The condition for a tie-chain is that two stem vectors, which are not anti-parallel, are within the same chain. More precisely, the second stem continues in some other direction and enters a second lamella instead of folding back on itself and reentering the same lamella in the anti-parallel scenario.  

\subsection{Entanglements Analysis}

Entanglements are analysed using the method introduced by Everaers et al \cite{EvSuGrEA2004sci,SuGrKrEv2005jpsb}. It consists in a chain contraction via energy minimization after fixing the chain ends, switching off intra-molecular excluded volume, and setting the equilibrium bond lengths to 0. For the branched systems, we removed the short chain branches as they are too short to be topologically relevant. All chains are contracted simultaneously, but the analysis is then done separately for long and short chains of the bidisperse system.

\section{Results and Discussion}
\label{sec:results}

\begin{figure}[htb!]
\includegraphics[width=0.85\textwidth]{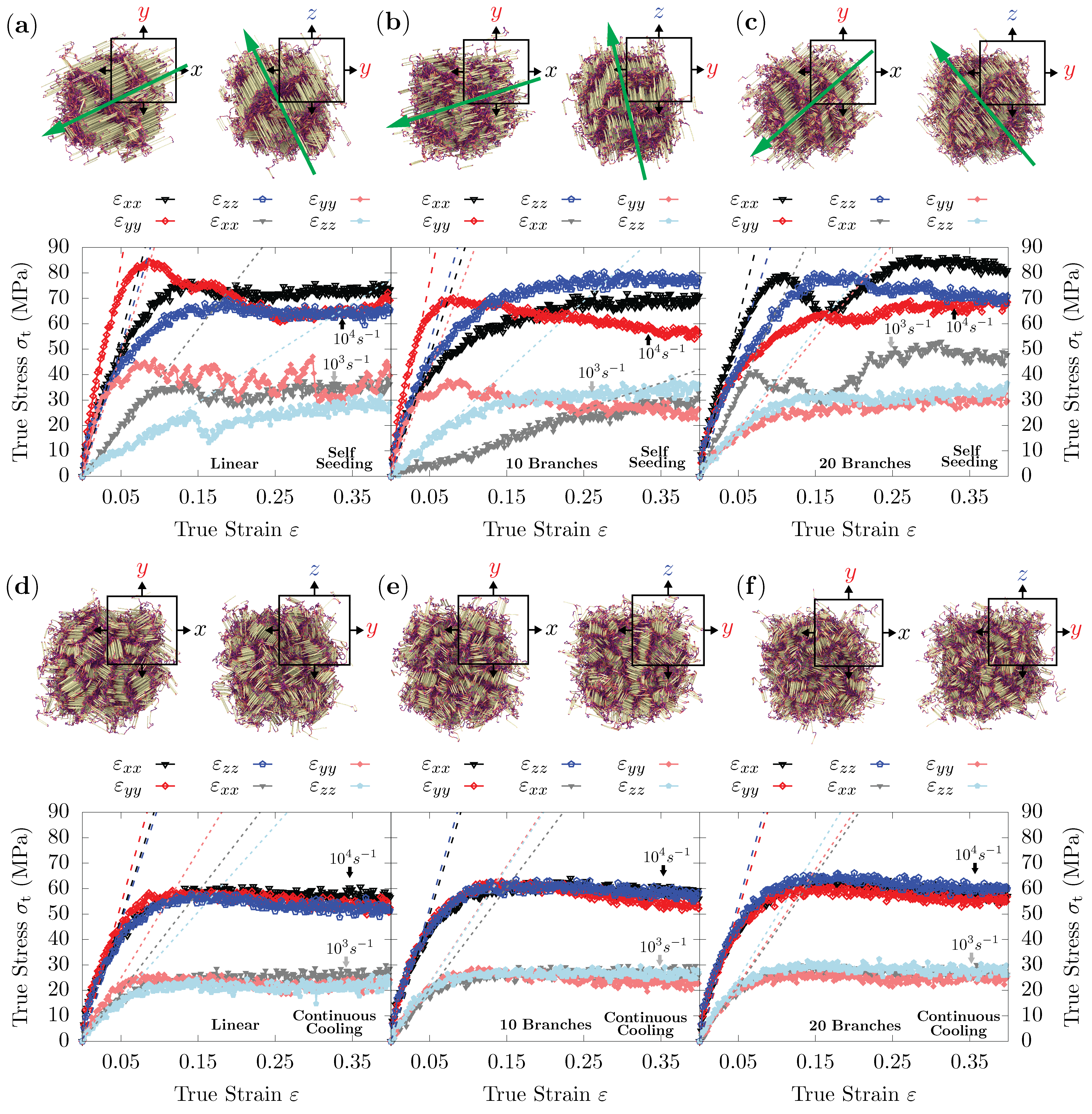}
\caption{Comparison of small strain behaviour between single and multi-domain crystals, grown via self-seeding or continuous cooling respectively, during uni-axial deformation in $x,y$ and $z$ directions. Panels (a-c) show stress-strain curves of single PE crystals with either, 0, 10 and 20 branches, for two different deformation rates $10^{4}s^{-1}$ (fast) and $10^{3}s^{-1}$ (slow) as distinguished by the regular and pastel colours. Snapshots show the orientation of the crystal w.r.t the simulation box and the green arrow indicates the global nematic director of the (beige) crystalline stems.  Panels (d-f) show similarly stress-strain curves during uni-axial deformation for their multi-domain counterparts grown via continuous-cooling. Note all curves overlap strongly in panels (d-f) demonstrating all directions are near mechanically equivalent. Dashed lines in all panels are a linear fit of the Young modulus at small strains.}
\label{fig:youngs}
\end{figure}

The directional dependence of the elasticity of each of the different systems can first be addressed by examining the stress-strain curves at small strains as shown in Fig. \ref{fig:youngs}. Beginning with the self-seeded systems, panel (a) shows the beginning of the stress-strain curves up to a true strain of $\varepsilon=0.4$ in the $x$ (black), $y$ (red) and $z$ (blue) directions respectively for the linear unbranched system. The solid and pastel colours distinguish the fast and slow rates respectively. A pronounced impact of the direction of deformation is immediately apparent for both deformation rates considered. The $x$ and $z$ directions show similar behaviour, whilst the $y$ direction is markedly different. This arises due to the orientation of the crystalline domain in the simulation cell, as shown in the snapshot at the top of panel (a). In the $x$ and $z$ directions the global crystalline stem direction (nematic director) is near parallel to the direction of deformation whereas in the $y$-direction it is near perpendicular. In the elastic regime, this results in a smaller Young modulus in the $x$ and $z$ directions (black and blue curves), while for larger strains the stress-strain curves gradually cross over to a plateau after a shallow peak. In comparison, the $y$ direction (red curve) is characterized by a larger Young modulus and shows a relatively sharper peak stress with a pronounced yield point at smaller strains, indicating that deformation near perpendicular to the global stem direction results in a faster transition to irreversible plastic behavior. This is consistent with experimental observations which demonstrate the modulus of flow oriented polymer films is lower in the flow direction (parallel) compared to cross flow (perpendicular)\cite{schrauwen2004structure}. 

Similar behaviour is observed in the system with 10 SCBs in panel (b), although it is more pronounced due to the crystalline stem direction aligning near perfectly perpendicular to the $y$ direction of the simulation cell. The $x$ and $z$ directions (black and blue curves) behave similarly with a shallow peak and gradual plateau, whereas the $y$ direction (red curve) has a more pronounced yield point (peak) and steeper incline indicating a higher Young modulus, as in the case of the linear system. In panel (c) the system with 20 SCBs appears special due to the crystalline stem direction aligning along the (111) direction of the box at a 45$^\circ$ angle to the principal axes. All directions are therefore near equivalent in terms of morphology. However, differences are still visible. The $y$ and $z$ directions (red and blue curves) show similar responses to deformation parallel to the global stem direction seen previously in panels (a) and (b). The $x$ direction (black curve) behaves differently. In the elastic regime, the Young modulus is larger than for $y$ and $z$ directions, indicating that deformation takes place primarily perpendicular to the stem direction. Moreover, after the yield point, the stress-strain curves passes through a minimum, beyond which strain hardening occurs before the stress appears to plateau for $\varepsilon > 0.25$.

In contrast to the self-seeded systems, the equivalent multi-domain systems grown by continuous-cooling in panels (d-f) show no direction dependence with all curves overlapping for a given rate of deformation. This preparation protocol therefore leads to near isotropic behaviour. The peak stress at the yield point between systems appears to increase marginally with increasing SCB content but this trend is weak. On this basis, we conclude short chain branching plays little or no role in mechanical response at small strains. 

\begin{figure}[htb!]
\includegraphics[width=0.85\textwidth]{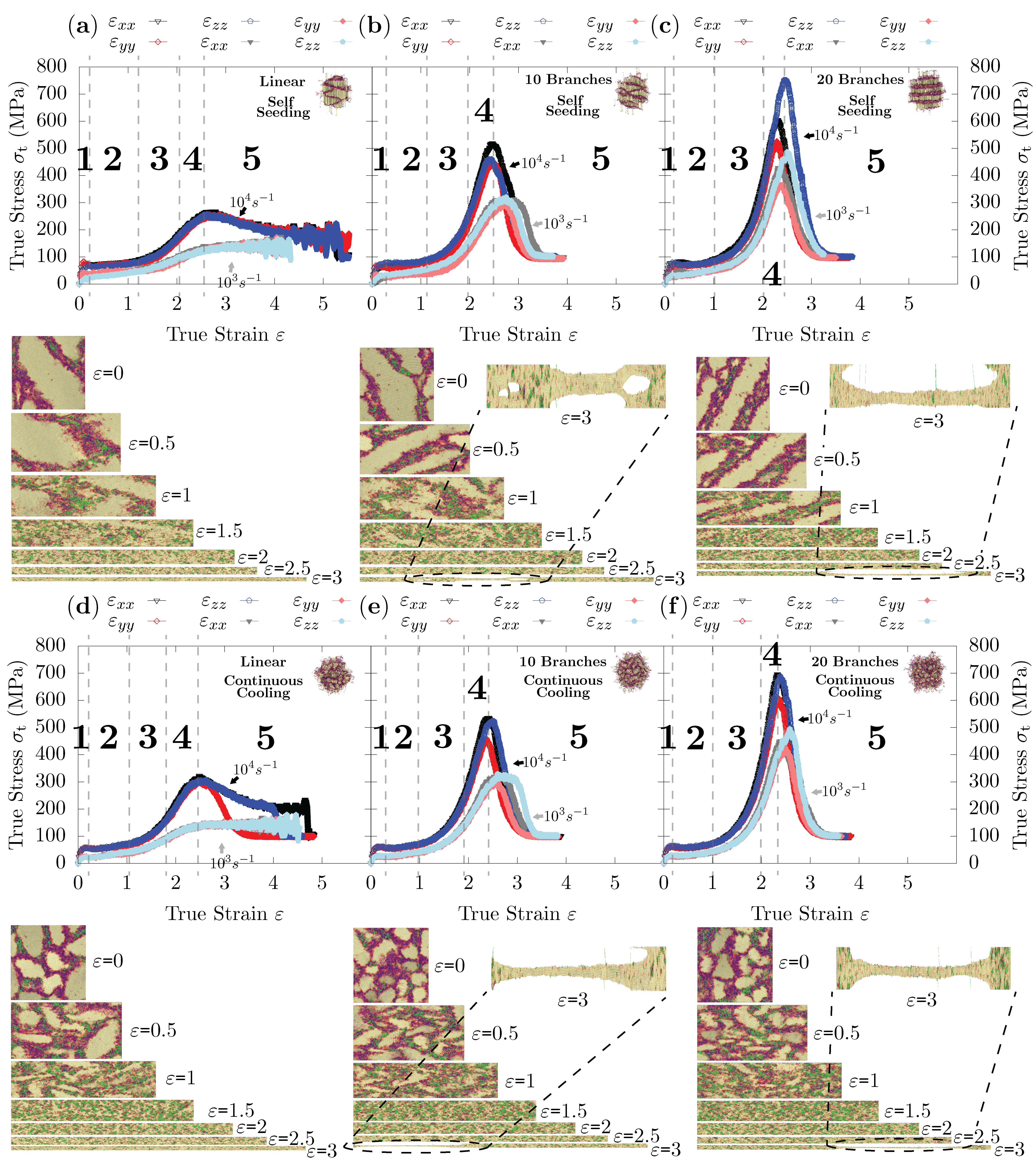}
\caption{Large strain behaviour comparison between single and multi-domain crystals, grown via self-seeding or continuous cooling respectively, during uni-axial deformation in $x,y$ and $z$ directions.  Panels (a-c) show stress-strain curves of single PE crystals with either, 0, 10 and 20 branches, for two different deformation rates $10^{4}s^{-1}$ (fast) and $10^{3}s^{-1}$ (slow) as distinguished by the solid and pastel colours. Selected snapshots below show the evolution of the amorphous (purple) and crystalline (beige) regions during the tensile test and the appearance of craze. Panels (d-f) show similarly stress-strain curves during uni-axial deformation for their multi-domain counterparts grown via continuous-cooling. Dashed lines in all panels indicate the 5 distinct regions of the stress-strain curve corresponding to pre-yield, post-yield plateau and strain hardening, strain hardening, strain-hardening and failure and post failure.} 
\label{fig:mechanical}
\end{figure}  

At larger strains, on the other hand, the effect of short chain branching has a strong influence on mechanical properties, as shown in Fig.~\ref{fig:mechanical}. Panel (a) shows the full stress-strain curve for the linear single-domain system where black, red and blue curves correspond to deformation in the $x$, $y$ and $z$-directions of the simulation cell, as defined previously. The solid and pastel colours distinguish between the fast and slow rates, respectively. Each curve is demarcated into 5 distinct regions corresponding to (1) pre-plateau region containing the Young modulus (2) post-yield plateau and strain hardening, (3) strain hardening, (4) strain-hardening and failure and (5) post failure. The shape of the stress-strain curve in panel (a) suggests the direction dependence seen in the Young modulus region (1) is no longer present at large strains, where all curves overlap with one another. This is likely due to destruction of the multi-lamellar crystals during the tensile test. The snapshot series below panel (a) illustrates this more clearly, where the chains backbones are coloured according to their local crystallinity from crystalline (yellow) to amorphous (purple) and tie-chain segments are highlighted in bright green. At the plateau of the stress-strain curve in the post-yield/strain-hardening region (2), the corresponding snapshot ($\varepsilon = 1$) suggests that the initial crystal structure is reoriented and partially deformed (squashed). However, in the strain-hardening region (3) the multi-lamellar structure can be seen to disappear in its entirety where it instead resembles a near homogeneous mixture of small amorphous and crystalline regions ($\varepsilon =2$). The significant rearrangement and restructuring of the chains is directly related to change in the gradient of the stress-strain curve. In the failure region (4), the peak-stress indicates the beginning of craze, where large cavities begin to appear. The peak appears broad which suggests the polymer is soft and easily deformed consistent with the snapshots below the stress-strain curves. The maximum strain in the linear system is particularly extreme, in the post-failure region (5), where the simulation cell reaches over 150 times its original length before the polymer completely breaks. Linear unbranched polymer chains therefore result in a highly ductile materials. 

We note this behaviour was not observed in similar computational studies for bidisperse linear chains which otherwise suggest such systems are brittle, with a narrow failure region, contrary to our findings \cite{moyassari2019molecular,moyassari2019molecular2}. This could be due to the higher molecular weight chains and low temperature, slow deformation used in our study, in comparison to the high temperature fast deformation favoured in the aforementioned study. It is notable that the speed of deformation has a pronounced effect on the failure region, where the faster the deformation the sharper the peak at failure. Typical strain rates used in experiments for polymers vary between 0.01-10 s$^{-1}$, thus the rapid timescales of simulation cannot be overlooked. The consideration of multiple (slow) rates and extrapolation to long times is therefore necessary when drawing conclusions from tensile tests in simulations. \Will{The deformation rates considered in this study are already at the limit of what is computationally affordable, where the slowest deformation rate considered in the linear system took approximately 10 days per direction on 640 cores of Jean-Zay.} In the next section, the morphological features responsible for material properties i.e. tie-chains or entanglements will be discussed in more detail. 

The introduction of 10 SCBs, in panel (b), results in a similar picture at the plateau of the stress-strain curve in the post-yield/strain-hardening region (2) where the corresponding snapshot ($\varepsilon = 1$) shows the original multi-lamellar structure is still visible. In the strain-hardening region (3) the multi-lamellar structure then disappears, resembling a near homogeneous mixture of small amorphous and crystalline regions ($\varepsilon = 2$). Crucially, this takes place over a much shorter strain region and is much steeper than for the linear system, suggesting SCBs prevent the relaxation and rearrangement of chains. The peak stress at failure is much higher than that of the linear system in region (4) and marks the beginning of craze and cavitation, see the snapshot corresponding to $\varepsilon =3$. This occurs at similar strains to the linear system but the strain range over which the material fails is drastically shorter where the simulation box reaches only 20 times its original length before a complete brittle break occurs. This suggests the introduction of SCBs results in a material with a greater tensile strength, which is less ductile. We note that some directional dependence is also apparent, but this is weak given the original multi-lamellar crystal is destroyed at large strains. 

At very high SCB content, of 20 SCBs per HMW chain, the trend is similar to the 10 SCB system, where the rise in region (3) is even steeper and the stress-strain peak is sharper and narrower for bother rates considered. Some direction dependence can be seen in the maximum peak-stress but this is not consistent between different rates. The peak stress of the slowest rate with 20 SCBs is comparable to the fastest rate with 10 SCBs in this instance. Furthermore, the maximum strain is further reduced in comparison to the 10 SCB system in panel (b) where all systems exhibit a complete brittle break before the box reaches 20 times its original length in region (3). The addition of a few SCBs is therefore sufficient to shift mechanical properties somewhere between that of ductile and brittle materials, giving bimodal branches PE systems their unique properties \cite{krishnaswamy2008effect}. 

This behaviour is similar across the multi-domain systems which are almost identical to their single-domain counterparts with the exception of the linear system. Remarkably, the multi-domain system exhibits clear direction dependence with 3 very different stress strain curves in $x$, $y$ and $z$ for the fastest rate considered. This is brought on by the more complicated (less ordered) topology and a greater number of entanglements which will be analysed in the proceeding section. The directional dependence is completely lost at slower deformation rates, where the chains have sufficient time to disentangle.

\begin{figure}[htb!]
\includegraphics[width=0.85\textwidth]{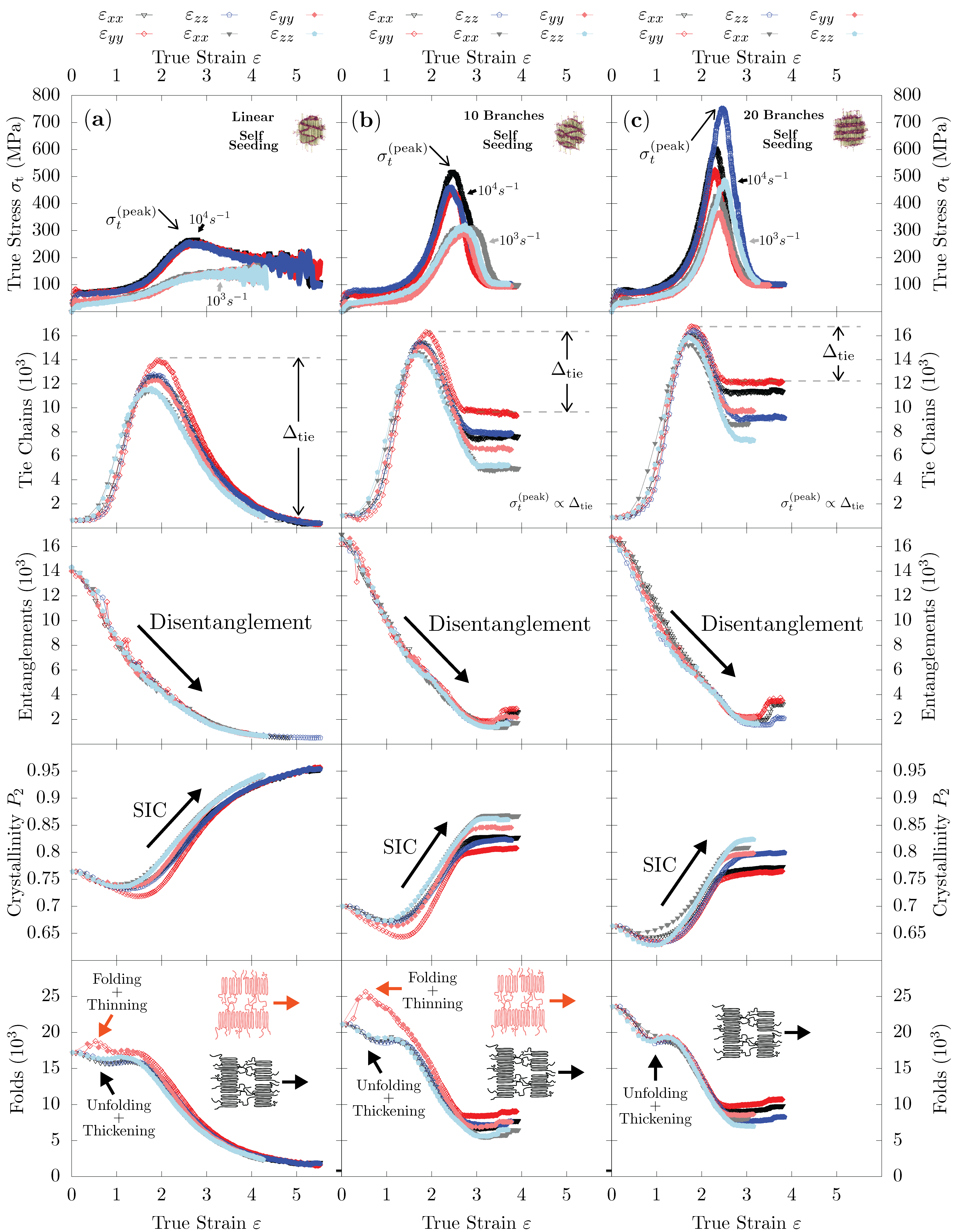}
\caption{True stress, tie segments, entanglements, crystallinity and folds vs true strain during tensile testing for single PE crystals with either 0, 10 or 20 branches, panels (a-c) respectively. The two different deformation rates $10^{4}s^{-1}$ (fast) and $10^{3}s^{-1}$ (slow) are distinguished by the solid and pastel colours and inset arrows. Note $\sigma_{\mathrm{t}}^{(\mathrm{peak})}$, $\Delta_{\mathrm{tie}}$ and SIC correspond to peak true stress, the change in number of tie chains from their peak to plateau after failure and strain induced crystallisation.}
\label{fig:statistics}
\end{figure}  

Figure \ref{fig:statistics} (a-c) show the stress-strain curves of the single-domain systems alongside the evolution of the number of tie-chains segments, entanglements, the overall crystallinity and number of folds. This allows direct links between morphological features and mechanical properties to be identified. Beginning with the linear system in panel (a), there seems to be little in the way of identifiable trends between the number of tie-chains or entanglements and the stress-strain behaviour. The number of tie-chains reaches a maximum in the strain-hardening region (3) suggesting multiple interconnected domains. This is best reflected in the snapshots in Fig.~\ref{fig:mechanical} (a) where the tie-chains segments are highlighted in green. It can be seen on destruction of the multi-lamellar crystal, that the amorphous (purple) and crystalline (yellow) regions appear almost homogeneously mixed. This is consistent with the initial decrease in crystallinity as the multi-lamellar crystal is first destroyed and then the subsequent increase as strain induced crystallisation (SIC) begins. The number of entanglements, on the other hand, appears to gradually reduce in number, suggesting the chains are disentangling as the polymer is deformed. This is also apparent in the number of folds, which gradually reduce consistent with disentanglement and SIC, where the polymer forms fibrils and becomes almost perfectly crystalline. We note also at small strains, deformation perpendicular to the crystalline stem direction results in an initial increase in folds, suggesting lamellar thinning, whereas deformation parallel shows only a simple decrease in the number of folds and lamellar thickening.

The introduction of 10 or 20 SCBs, in panels (b) and (c), changes the behaviour dramatically, where pronounced plateaus can be seen in both the number of tie-chains and entanglements. The number of tie-chains follows closely the stress-strain curve with an initial peak followed by a plateau in the number of tie-chain segments. Crucially, this behaviour is not seen in the linear system discussed previously, which shows only a gradual continuous decrease with increasing strain. The crystallinity is also substantially lower than that of the linear system which suggests SCBs prevent impede strain induced crystallisation and disentanglement of the chains. Such behaviour is not unexpected since SCBs are known to be forced into the amorphous region upon crystallisation \cite{fall2022role,fall2023molecular}. This further explains why the number of tie-chains and entanglements do not simply gradually decrease with increasing strain. The number of tie-chain segments in particular are clearly correlated with the stress-strain curve, which leads us to define the quantity $\Delta_{\mathrm{tie}}$ as the change in the number of tie-chains from the beginning of the brittle break (peak in tie-chains) and complete failure where the stress-strain curve reaches a maximum ($\sigma_{t}^{(\mathrm{peak})}$) and the tie-chains reach a plateau, see Fig.~\ref{fig:statistics}. The change in the number of tie-chains $\Delta_{\mathrm{tie}}$ appears proportional to the peak-stress $\sigma_{t}^{(\mathrm{peak})}$ which indicates tie-chain segments are driving the mechanical response in branched PE systems and highlights their importance. Entanglements, on the other hand, show no such trend, suggesting they play a secondary role in the mechanical response, at least in branched PE. The small differences between deformation directions, can be attributed to the change in the number of tie-chain segments $\Delta_{\mathrm{tie}}$, which differ depending on the chosen direction. This relationship always remains correlated with the peak stress in branched systems and will be analysed in more detail later. There is little noticeable difference between the tie-chain or entanglement curves with either 10 or 20 SCBs. We note, however, that the spread in $\Delta_{\mathrm{tie}}$ is narrower with increasing SCB content. This suggests chain sliding and disentanglement is further suppressed with a greater number of SCBs and that the tie-chain segments feel the deformation more strongly as a result and dominate the mechanical response almost entirely. 

\begin{figure}[htb!]
\includegraphics[width=0.85\textwidth]{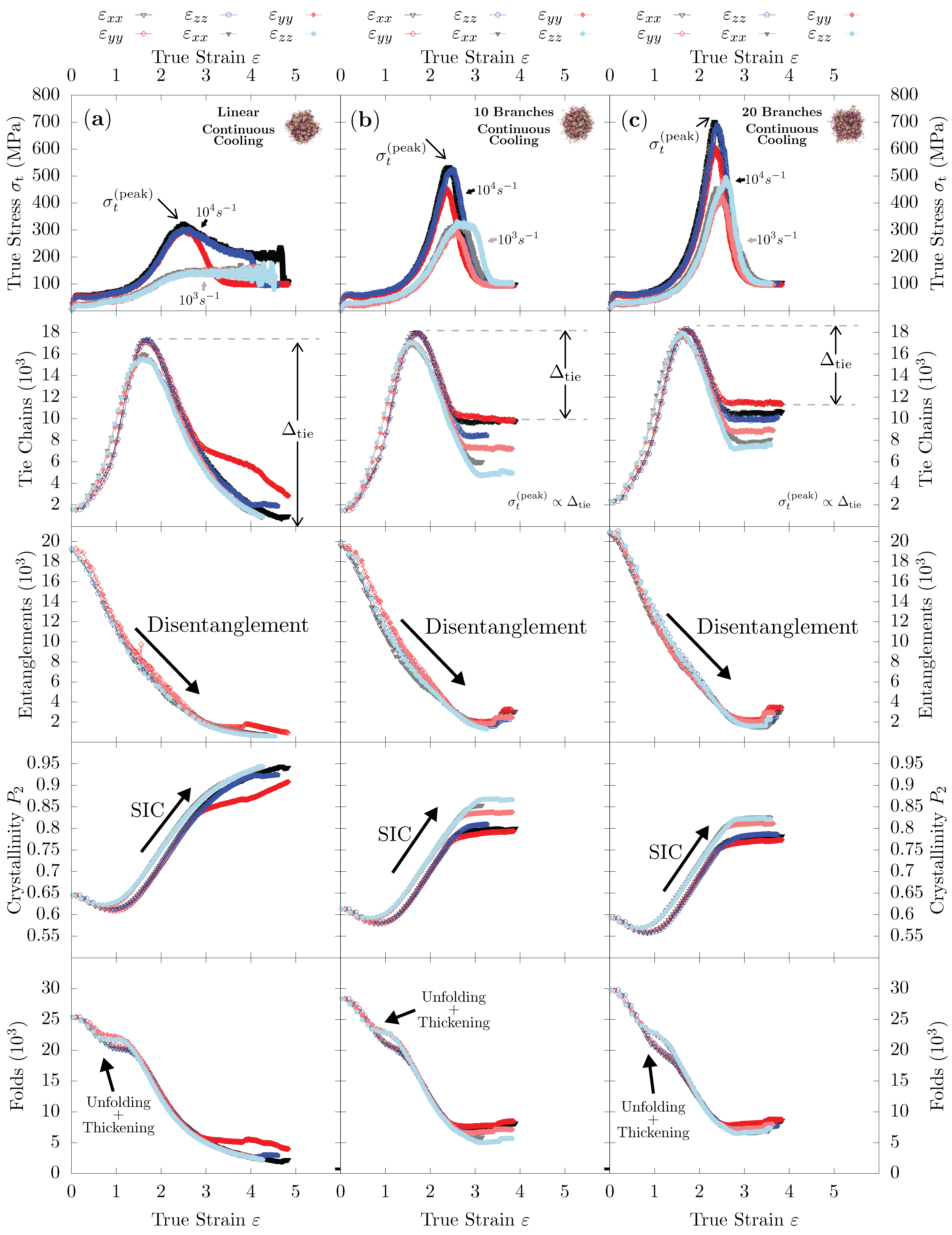}
\caption{True stress, tie segments, entanglements, crystallinity and folds vs true strain during tensile testing for multi-domain PE crystals with either 0, 10 or 20 branches, panels (a-c) respectively. The two different deformation rates $10^{4}s^{-1}$ (fast) and $10^{3}s^{-1}$ (slow) are distinguished by the solid and pastel colours and inset arrows. Note $\sigma_{\mathrm{t}}^{(\mathrm{peak})}$, $\Delta_{\mathrm{tie}}$ and SIC correspond to peak true stress, the change in number of tie chains from their peak to plateau after failure and strain induced crystallisation.}
\label{fig:statistics2}
\end{figure}  

The multi-domain systems in Fig.~\ref{fig:statistics2} show broadly similar behaviour despite the very different initial morpholgies pre-deformation. This further supports our conclusions that pre-oriented multi-lamellar PE crystals are mostly destroyed outside region (1) of the stress-strain curve where deformation is permanent. Much of the initial memory is therefore lost, suggesting initial global chain alignment plays little role in the mechanical response at large strains. We do however note one exception, as mentioned previously, where the linear system exhibits clear direction dependence only for the fastest deformation rate considered. \Will{The x and z directions (blue and black curves) in Fig \ref{fig:statistics2} (a) show that the number of tie-chains reach a small plateau instead of a the gradual decrease expected. In addition the y-direction shows a more pronounced plateau with one further additional plateau present in the entanglements (red curve). We note from the small strain behaviour shown in Fig \ref{fig:youngs} (d) all directions appear mechanically equivalent in multi-domain systems thus it is surprising that direction dependence is showing up under fast deformation at large strains. We speculate that the rapid deformation prevents the reorientation of crystalline domains and that this in turn impedes reorganisation and disentanglement of the chains. Whilst this effect is clearly dependent on both the deformation rate and the finite size of the simulation box, it is nonetheless revealing that the multi-domain topology plays a special role in the mechanical response}
 
\begin{figure}[htb!]
\includegraphics[width=0.85\textwidth]{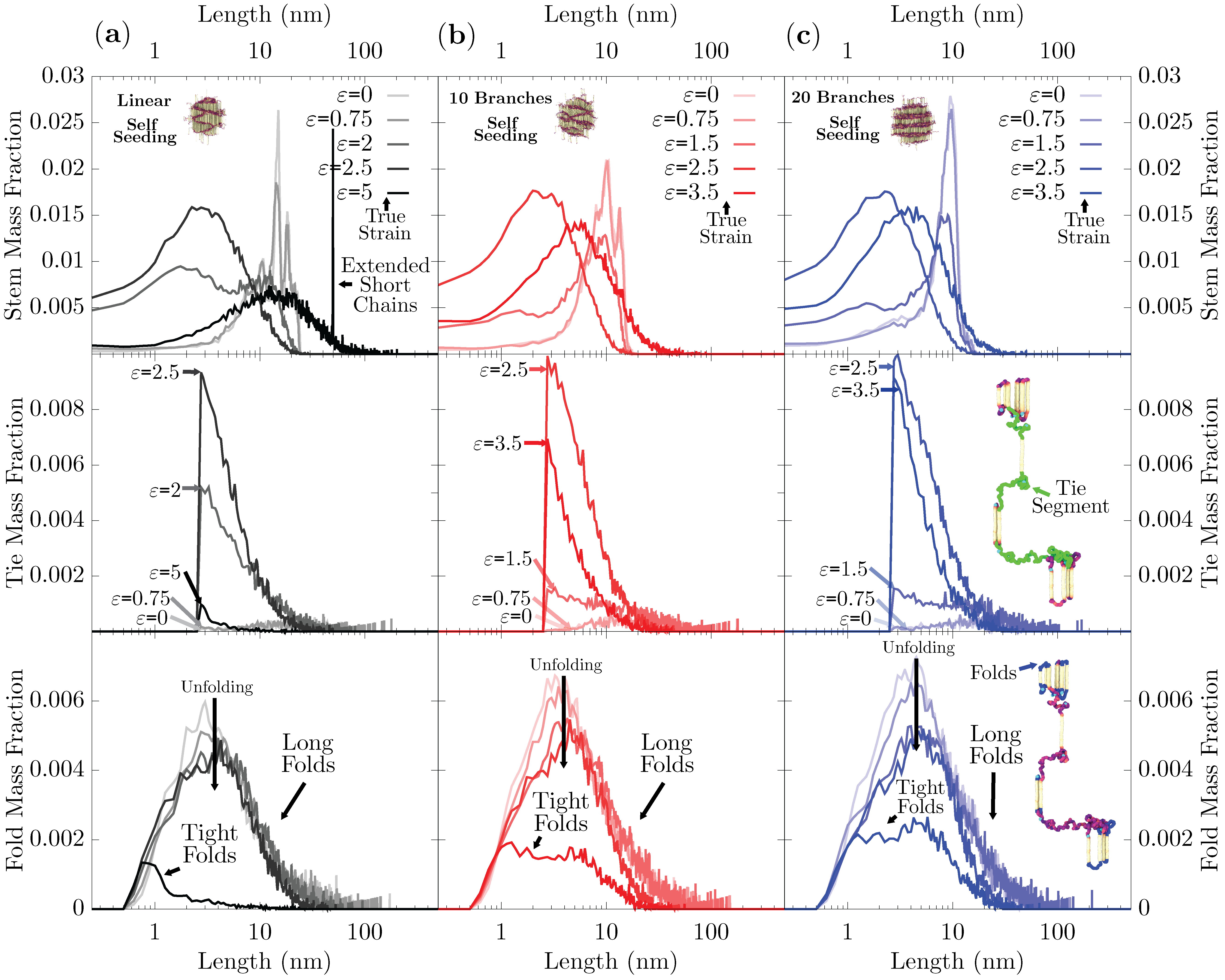}
\caption{Crystal stem, tie chain segment and fold segment length distributions for all multi-domain PE crystals with either 0, 10 or 20 branches, panels (a-c) respectively. The increasing true strain is indicated by the lighter to darker coloured curves and true strain is indicated in both the key and with inset arrows in the tie segment distributions. Note the sharp peak in row 1 panel (a) corresponds to a large fraction of fully extended short 200 united-monomer long chains present at large strains. This feature is missing in systems including short chain branches \Will{which prevent the chains from fully extending.} Such trends are common traits in all systems, both multi and single domain, other distributions can be found in Supplementary Information.}
\label{fig:distributions}
\end{figure}  

In Fig.~\ref{fig:distributions}, the distribution of the lengths of crystalline stem-lengths, tie-chain segments and folds shown for the single domain systems only in the $x$-direction. We note that trends between systems are broadly similar, hence only a selection is shown here. The distributions are taken at different points along the stress-strain curve where lighter colours correspond to smaller strains as indicated in the figure. The light grey curve corresponding to $\varepsilon = 0$ shows the initial distribution of crystalline stem lengths as reported previously \cite{fall2023molecular}. With increasing strain, the mass fraction of long stems appears to reduce ($\varepsilon = 0.75$) but the shape of the multi-peaked distribution does not change, consistent with the preservation of the initial crystal in the region (1), where deformation is not permanent, see the snapshots in Fig.~\ref{fig:mechanical} (a). At larger strains, the distribution resembles that of a Gaussian and all sharp peaks of the initial multi-lamellar crystal are lost around $\varepsilon = 2$. This is consistent with the destruction of the initial topology at large strains and explains why little difference in mechanical response can be seen between single or multi-domain systems. At very large deformation ($\varepsilon =5$) a noticeable spike is observed for stems corresponding to 200 united-monomers. This results from a large fraction of the short (LMW) chains fully extending due to SIC. A large proportion of short chains therefore fully disentangle in linear systems. No such peak is observed in the distributions corresponding to the systems with either 10 or 20 SCBs, which supports our conclusions that SCBs prevent disentanglement of the LMW chains. 

The distribution of tie-chain segments shows interesting characteristics where clear differences arise between linear and branched systems. There are relatively few tie-chain segments at small strains in all systems but we note that there are more in the branched systems in comparison to the linear ones. This is due to the thinner lamellae induced by short chain branching, which allows the shorter (LMW) chains to form more tie-chains between thinner lamellae. In the linear system, upon deformation, the tie segment length distributions resemble a log normal distribution, exhibiting a peak at approximately 2.5 nm, which corresponds to a minimum possible tie segment length of 10 united monomers for a tie-chain segment to cross the amorphous region between crystalline domains. A long tail exists which in the initial topology may extend as far as 400 united monomers, which is extremely long for a tie-chain segment. A sample of tie-chains segment conformations are shown in panel (c) where the crystalline and amorphous regions are coloured yellow and purple respectively with tie-chain segments shown in green. On deformation, the number of tie-chains increases and the tail of the distribution becomes shorter indicating tie-chain segments are more numerous and generally shorter above $\varepsilon = 0.75$. This is consistent with the destruction of the initial topology and near homogeneous mixture of amorphous and crystalline domains seen in the snapshots in Fig.~\ref{fig:mechanical} (a).  After the brittle break ($\varepsilon = 5$) the tie-chain segments are very small in both number and in length and this is due to the near perfect crystallinity of the remaining sample after SIC and full extension of a large proportion of the shorter chains, see Fig.~\ref{fig:statistics} (a) and is consistent with disentanglement and chains sliding.

The addition of SCBs does not appear to change the behaviour of the crystalline stem length distributions in panels (b) and (c) which again retain their original peaks until the multi-lamellar crystal is destroyed above $\varepsilon =2$. The distributions behave similarly until the tie-chain segments reach their peak mass fraction, after which the tie-chain segments persist in the branched systems with higher maxima at equivalent strains and a persistent long tail. The mass fraction of tie-chains remains high after the brittle break in branched systems in contrast to the linear system which is almost entirely devoid of tie-chains. This is again consistent with chain disentanglement, SIC and extension of the short chain component in the linear system which is not present in the branched systems, suggesting SCBs cause tie-chain segments to persist at large strains in branched systems. It is interesting to note that the mass fraction of tie-chain segments increases marginally with increasing SCB content. The fold length distributions, on the other hand, appear broad and gradually reduce as the chains unfold, as reflected in Figs.~\ref{fig:statistics} and \ref{fig:statistics2}. In the linear system, the distribution of fold lengths remains similar until after the brittle break where the center of the fold length distribution shifts to small values with a large reduction in peak height, indicating only a small mass fraction short folds remain. In the branched systems, the fold length distributions behave similarly. However, the distributions remain broad after the brittle break, suggesting folds still account for a significant mass fraction of the system and that the semi-crystalline morphology is persisting with SCBs with a significant spread of fold lengths. 

\begin{figure}[htb!]
\includegraphics[width=0.85\textwidth]{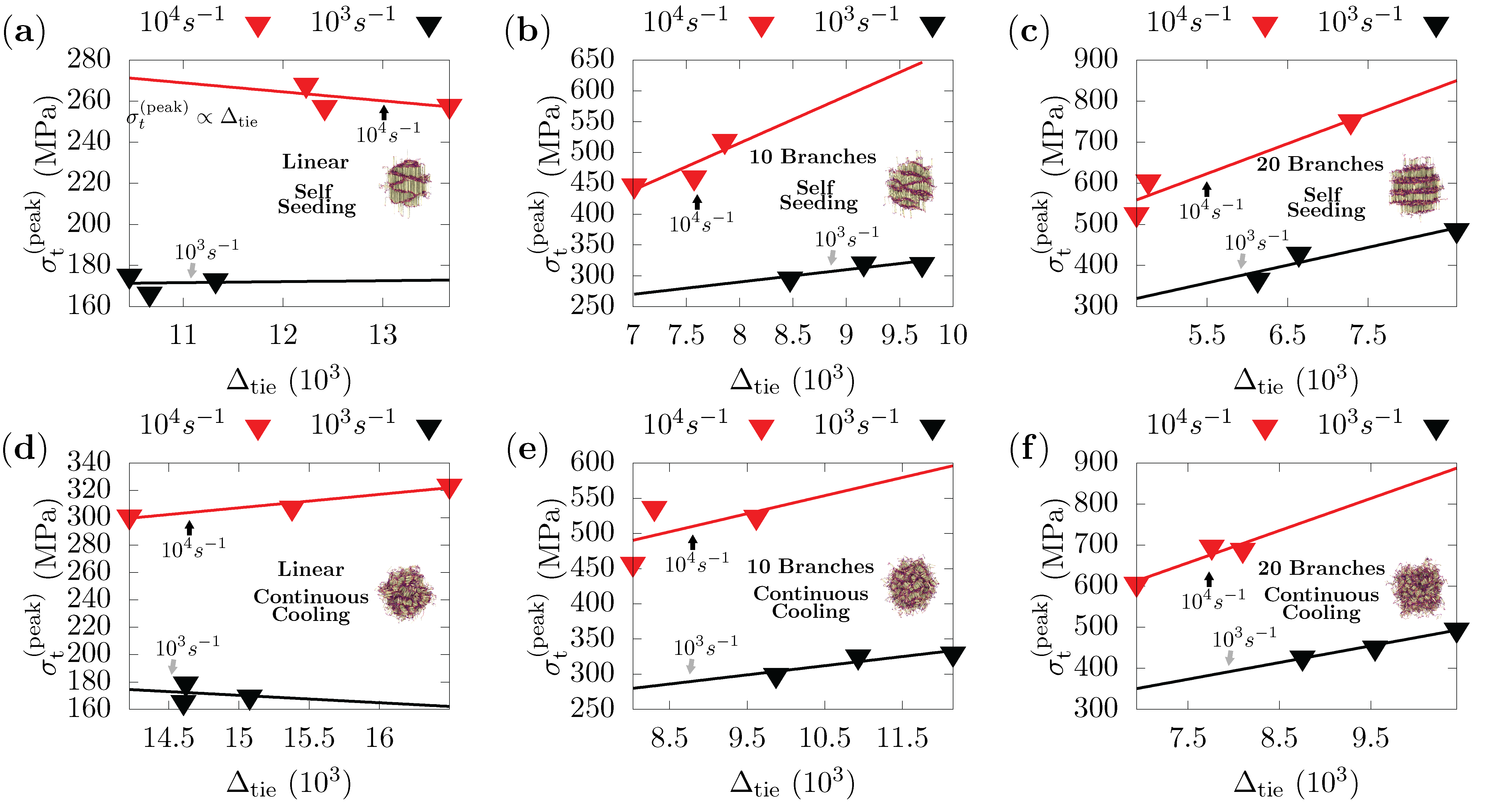}
\caption{Peak stress vs change in tie-chain segments for single (a-c) and multi-domain (d-f) systems. The red and black points correspond to fast and slow deformation rates respectively and the lines are a linear fitting.}
\label{fig:trends}
\end{figure}  

In Fig.~\ref{fig:trends} the relationship between $\Delta_{\mathrm{tie}}$ and the peak-stress $\sigma_\mathrm{t}^{(\mathrm{peak})}$ is examined more closely for all systems in all directions at all rates considered. Panels (a-c) and (d-f) correspond to single and multi-domain systems respectively and show $\Delta_{\mathrm{tie}}$ vs $\sigma_{t}^{(\mathrm{peak})}$ profiles for each of the 3 directions at a given rate. Fast and slow rates correspond to red and black curves respectively where each curve contains 3 data points, one for each direction of deformation. In both linear systems, panels (a) and (d) we note that the relationship $\sigma_\text{t}^{(\mathrm{peak})} \propto \Delta_{\mathrm{tie}}$ 
does not hold in general, which supports our conclusions that tie-chains do not dominate the mechanical response in linear systems. On the other hand, in the branched systems, both multi and single domain at all rates considered, the relationship appears to hold with all curves showing a clear upward rise $\sigma_{t}^{(\mathrm{peak})}$ with increasing $\Delta_{\mathrm{tie}}$. This suggests that tie-chains dominate the mechanical response in branched systems and that the larger the number of tie-chain segments directly involved in the brittle break, the higher the peak stress or the greater the tensile strength of the material. 

\section{Conclusions}
\label{sec:conclusions}

Using the united monomer model of PE reported in Ref.~\citenum{fall2022role} and the subsequent multi-lamellar crystals grown in Ref.~\citenum{fall2023molecular}, the mechanical properties of bimodal PE systems have been tested using non-equilibrium molecular dynamics simulations to mimic uni-axial tensile testing. The well-aligned single-domain PE crystals allowed us to test the directional dependence on the Young's modulus behaviour at small strains and mechanical response at much larger strains. Using newly developed analysis tools, the number of tie-chain segments, entanglements, crystalline stems and folds could be tracked along the stress-strain curves at two different rates, allowing us to make the link between molecular morphology and mechanical behaviour. Our major conclusions include: 
\begin{enumerate}
    \item The introduction of SCBs regulates the crystalline stem length resulting in thinner lamellae, allowing the short chain component to form a greater number of tie-chains and entanglements. Multi-domain systems have more entanglements and tie-chains than their single-domain counterparts. 
    \item At small strains, deformation parallel to the global stem direction results in a smaller Young modulus
    than deformation perpendicular.
    \item Directional dependence and memory of the initial topology on the mechanical response is entirely lost at large strains where existing structures within the semi-crystalline polymer are destroyed.
    \item \Will{Faster deformation rates may impede the reorientation of crystalline domains in multi-domain systems preventing the removal of tie-chains and/or entanglements and the appearance of a less ductile (more brittle) material. }
    \item \Will{Linear unbranched chains exhibit poorer mechanical properties due to chain sliding, leading to a near complete loss of tie chains and chain entanglement, accompanied by substantial SIC.}
    \item Branched systems prevent chain sliding, complete disentanglement and substantial SIC. This ensures the semi-crystalline morphology persists at larger strains, resulting in large numbers of tie-chains segments between crystalline domains. Tie chains therefore dominate the mechanical response in branched PE systems.
    \item In branched systems, the change in the number of tie-chains during the brittle break is proportional to the peak stress at failure of the material. Entanglements appear to play a secondary role in the mechanical response of heavily branched PE systems where no clear relationship could be identified with stress-strain behaviour. 
\end{enumerate}

The findings presented here provide molecular level insight into the remarkable properties of bimodal branched polyethylene systems as well as PE systems in general. Our results suggest that the unique properties of bimodal branched PE arise from the reduced mobility of the chains, resulting from short chain branching, during deformation and the suppression of chain sliding, full disentanglement and substantial SIC. This leads to the persistence of tie-chain segments at large strains where the number of tie-chains involved in the brittle break directly relates to the tensile strength. Whilst we have addressed the role of morphological features on the mechanical response of these systems, the bimodal aspect of these systems and its role on the mechanical response is unclear. Since other molecular weight distributions have not been considered, including branches, it is difficult to draw any meaningful conclusions about the effect of the molecular weight distribution on the mechanical response. Work on branched systems with different molecular weight distributions is currently underway and will be published in a forthcoming study.





\begin{acknowledgement}
 We thank the High Performance Computing Center CAIUS of the University of Strasbourg for supporting this work by providing access to computing resources, partly funded by the Equipex Equip@Meso project (Programme Investissements d'Avenir) and the CPER Alsacalcul/Big Data. W.S.F wishes to thank Prof.\ Tom Engels and Prof.\ Leon Govaert for their helpful comments. W.S.F acknowledges HPC resources of IDRIS under the allocation 2023-A0150913823 made by GENCI.
\end{acknowledgement}

\begin{suppinfo}
Sample LAMMPS scripts and angular potentials used in this work. 
\end{suppinfo}

\bibliography{fall_et_al}

\end{document}